\documentclass[conference]{IEEEtran}
\IEEEoverridecommandlockouts
\usepackage{amsmath,amssymb,amsfonts}
\usepackage{subfigure}
\usepackage{graphicx}
\usepackage{textcomp}
\usepackage{hyperref} 
\usepackage{xcolor}
\usepackage{extramath}
\usepackage{pgfplots}
\usepackage{wrapfig}
\usetikzlibrary{pgfplots.groupplots}
\usepackage[style=ieee,backend=biber,doi=false,isbn=false,url=false]{biblatex}
\usepackage{siunitx}
\usepackage{pgfplots}
\pgfplotsset{compat=1.17}
\usepackage{lipsum}
\def\BibTeX{{\rm B\kern-.05em{\sc i\kern-.025em b}\kern-.08em
    T\kern-.1667em\lower.7ex\hbox{E}\kern-.125emX}}

\usepackage{algcompatible}
\usepackage{algorithm2e}
\RestyleAlgo{ruled}
\usepackage{subcaption}
\usepackage[compatibility]{tikztensor}
\usetikzlibrary{external}
\tikzexternalenable

\let\oldrho\rho
\let\oldpsi\psi
\renewcommand{\psi}{\bm{\oldpsi}}
\renewcommand{\rho}{\bm{\oldrho}}

\usepackage{etoolbox}
\AtBeginBibliography{\small} 

\begin{document}

% Title
\title{Tensor Train Quantum State Tomography using Compressed Sensing \thanks{This work was supported by the Flemish Government's AI Research Program and KU Leuven Internal Funds (iBOF/23/064, C14/22/096). Shakir Showkat Sofi, Charlotte Vermeylen, and Lieven De Lathauwer are affiliated with Leuven.AI - KU Leuven institute for AI, B-3000, Leuven, Belgium.}
}

% Authors
\author{\IEEEauthorblockN{Shakir Showkat Sofi}
\IEEEauthorblockA{\textit{Dept. Electrical Engineering (ESAT)} \\
\textit{KU Leuven}\\
Kortrijk, Belgium \\
shakirshowkat.sofi@kuleuven.be}
\and
\IEEEauthorblockN{Charlotte Vermeylen}
\IEEEauthorblockA{\textit{Dept. Electrical Engineering (ESAT)} \\
\textit{KU Leuven}\\
Leuven, Belgium \\
charlotte.vermeylen@kuleuven.be
}
\and
\IEEEauthorblockN{Lieven De Lathauwer}
\IEEEauthorblockA{\textit{Dept. Electrical Engineering (ESAT)} \\
\textit{KU Leuven}\\
Kortrijk/Leuven, Belgium  \\
lieven.delathauwer@kuleuven.be}
}

\maketitle

\begin{abstract}
Quantum state tomography (QST) is a fundamental technique for estimating the state of a quantum system from measured data and plays a crucial role in evaluating the performance of quantum devices. However, standard estimation methods become impractical due to the exponential growth of parameters in the state representation. In this work, we address this challenge by parameterizing the state using a low-rank block tensor train decomposition and demonstrate that our approach is both memory- and computationally efficient. This framework applies to a broad class of quantum states that can be well approximated by low-rank decompositions, including pure states, nearly pure states, and ground states of Hamiltonians. 
%To bridge the gap between quantum computing and signal processing, we present our method using signal processing terminology, aiming to inspire cross-disciplinary research and the application of shared techniques.
\end{abstract}

\begin{IEEEkeywords}
low-rank approximation, quantum state tomography, tensor completion, tensor train
\end{IEEEkeywords}

\section{Introduction}
Quantum computing has garnered significant interest in recent years due to emerging tools that enable the analysis of large-scale systems and optimization over high-dimensional spaces. The study of quantum systems is increasingly crucial in both academia and industry, with applications in quantum communication, sensing, and control. A quantum system's state is represented by a density matrix—a Hermitian, positive semidefinite (PSD) matrix with unit trace. Density matrices provide a unified framework for representing both pure and mixed states (probabilistic mixtures of pure states) \cite{paris2004qse, nielsen2010quantum}.

Determining the density matrix of a quantum system is essential yet challenging. Quantum state tomography (QST) is a fundamental tool for benchmarking and verifying quantum devices \cite{paris2004qse, nielsen2010quantum, cramer2010efficient, lanyon2017efficient}. QST estimates an unknown density matrix by measuring an ensemble of identically prepared quantum systems \cite{paris2004qse, nielsen2010quantum, cramer2010efficient, gross2010csqst, liu2012csqst}. It involves two main steps: (i) performing measurements on a large number of identically prepared quantum systems/circuits and (ii) post-processing the data to reconstruct the density matrix. However, as the system size grows, the number of parameters in the density matrix increases exponentially---an obstacle known as the “curse of dimensionality.”

In practice, many quantum states of interest exhibit special properties, such as being nearly pure or corresponding to ground states of local Hamiltonians \cite{gross2010csqst, liu2012csqst}. In such cases, the density matrix has a low rank. This allows for more efficient estimation, as fewer measurements may suffice for accurate recovery. By leveraging connections between low-rank matrix completion and QST, convex optimization methods have been proposed to solve semidefinite programs (SDPs) with nuclear norm penalization. These approaches enable density matrix reconstruction with fewer measurements than traditional tomography methods \cite{gross2010csqst, liu2011universal, liu2012csqst, wang2013qstvmc, candes2012mc, kalev2015quantum}. However, they implicitly use rank information while optimizing over the full matrix, limiting their scalability.

For large-scale SDPs, an efficient heuristic is the Burer-Monteiro factorization \cite{burer2003lrsdp}, which explicitly imposes a low-rank structure. This approach reformulates the problem non-convexly but significantly improves efficiency by optimizing only over low-rank factors. Inspired by this idea, several QST algorithms employing a Cholesky-like low-rank parametrization of the density matrix have been developed \cite{bhojanapalli2016lrsdp, kyrillidis2018provable, kim2023mifgd}.

Another line of research models the density matrix as a high-order tensor and employs tensor networks—contracted networks of low-order tensors—for compression, thereby mitigating the curse of dimensionality. Common tensor network representations include matrix product states/operators (MPS/MPO) or tensor trains (TT) \cite{verstraete2007MPS, verstraete2004MPO, oseledets2010tensortrain, oseledets2010approximation}, tree tensor networks, and projected entangled pair states \cite{orus2014practical}. QST has been shown to be feasible for certain large-scale quantum many-body systems whose states can be approximated by low-rank tensor networks, such as ground states, GHZ states, cluster states, and AKLT states \cite{cramer2010efficient, lanyon2017efficient, qin2024quantum, kuzmin2024qst, khoromskaia2018tensor}. The MPS-based QST approach has demonstrated both efficiency and scalability \cite{cramer2010efficient, lanyon2017efficient, kuzmin2024qst}. In \cite{qin2024quantum}, QST is generalized for MPOs, extending MPS methods from pure states to noisy mixed states \cite{verstraete2004MPO}. However, ensuring that the reconstructed density matrix satisfies necessary constraints (Hermitian, PSD, and unit trace) remains a challenge.

\emph{Contributions:} In this work, we adopt a block tensor train (Block-TT) \cite{dolgov2014computation, lee2015btt} parametrization for the density matrix. Unlike other methods, this approach inherently preserves positive semidefiniteness without requiring additional constraints---analogous to the Cholesky decomposition for matrices. More importantly, our approach also helps mitigate the curse of dimensionality for a large number of states. We compare our method with state-of-the-art methods and show that it is both memory- and computationally efficient, while achieving competitive accuracy in recovering mixed states. 

\subsection{Preliminaries and Notation}  
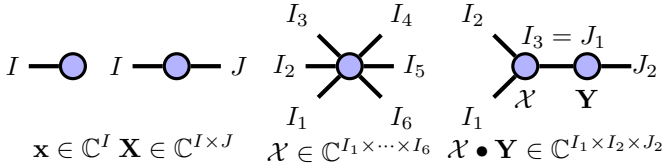
\begin{figure}[h]
	\centering
       \begin{tikzpicture}[bluenode/.style={circle, draw=black!100, fill=blue!30, very thick, minimum size=2mm}, scale=0.46]
\begin{scope}[xshift=-4cm, yshift=0]

% Vector
\node[bluenode] at (0, 0) (v0) {};
\node[left = 0.4 of v0] (v1) {$I$};
\node[right = 0.4 of v0] (v2) {};
\draw[ultra thick] (v0.west) -- (v1);

% Matrix
\node[bluenode] at (3, 0) (m0) {};
\node[left = 0.4 of m0] (m1) {$I$};
\node[right = 0.4 of m0] (m2) {$J$};
\draw[ultra thick] (m0.west) -- (m1);
\draw[ultra thick] (m0.east) -- (m2);

% Tensor
\node[bluenode] at (8, 0) (tm0) {};
\node[left = 0.4 of tm0] (tm1) {$I_2$};
\node[above left = 0.4 of tm0] (tm2) {$I_3$};
\node[above right = 0.4 of tm0] (tm3) {$I_4$};
\node[right = 0.4 of tm0] (tm4) {$I_5$};
\node[below right = 0.4 of tm0] (tm5) {$I_6$};
\node[below left = 0.4 of tm0] (tm6) {$I_1$};
\draw[ultra thick] (tm0.west) -- (tm1);
\draw[ultra thick] (tm0.north west) -- (tm2);
\draw[ultra thick] (tm0.north east) -- (tm3);
\draw[ultra thick] (tm0.east) -- (tm4);
\draw[ultra thick] (tm0.south east) -- (tm5);
\draw[ultra thick] (tm0.south west) -- (tm6);

% Contraction
\node[right = 1 of tm4, bluenode,label= south:{$\ten{x}$}] (ta0){};
\node[above left = 0.4 of ta0] (ta1) {$I_2$};
\node[below left = 0.4 of ta0] (ta2) {$I_1$};
\node[right = 0.2 of ta0, label= north:{$I_3 = J_1$}] (ta3) {};
\draw[ultra thick] (ta0.north west) -- (ta1);
\draw[ultra thick] (ta0.south west) -- (ta2);

\node[right = 0 of ta3, bluenode,label=south:{$\mat{y}$}] (tb0){};
\node[left = 0.4 of tb0, inner sep=0] (tb1) {};
\node[right = 0.4 of tb0, inner sep=0] (tb2) {$J_2$};
\draw[ultra thick] (tb0.east) -- (tb2);

\draw[ultra thick] (ta0.east) -- (tb0.west);

\coordinate (midpoint) at ($(ta0)!0.5!(tb0)$);
    
% Annotations
\node[below=0.6 of v0]{$\vec{x} \in \C^{I}$};
\node[below=0.6 of m0]{$\mat{x} \in \C^{I \times J}$};
\node[below=0.65 of tm0]{$\ten{x} \in \C^{I_1 \times \cdots \times I_6}$};

\node[below=0.8 of midpoint]{$\ten{x} \bullet \mat{y} \in \C^{I_1 \times I_2 \times J_2}$};

\end{scope}

\end{tikzpicture}
        \vspace{-5mm}
	\caption{\small Basic tensor diagrams.}
\label{fignotation}
\end{figure}

We use lower-case, bold lower-case, bold capital, and calligraphic letters to denote scalars, vectors, matrices, and tensors, respectively, i.e., $x, \vec{x}, \mat{x}, \text{ and } \ten{x},$ respectively. The {\em order} of a tensor $\ten{x} \in \C^{I_{1} \times I_{2} \times \cdots \times I_{N}}$ is the number of its {\em modes,} which corresponds to the number of free edges in the tensor network diagram, as shown in Fig.~\ref{fignotation}. The trace of a matrix $\mat{x}$ is denoted by $\operatorname{Tr}(\mat{x})$. The Frobenius norm and the nuclear norm (trace norm) are denoted by \(\|\cdot\|_{\mathrm{F}}\) and \(\|\cdot\|_*\), respectively. A density matrix is denoted by \(\rho\), and a Hilbert space by \(\mathcal{H}\). The outer product $\ten{Z} = \mat{X} \op \mat{Y}$ of two matrices $\mat{X} \in \C^{I_1 \times I_2}$ and $\mat{Y} \in \C^{J_1 \times J_2}$ satisfies $z_{ijkl} = x_{ij} y_{kl}$ and the Kronecker product $\mat{Z} = \mat{X} \otimes \mat{Y}$ of the same matrices satisfies $z_{k+(i-1)J_1, l+(j-1)J_2} = x_{ij} y_{kl}$. A product (contraction) between the last mode of tensor $\ten{x} \in \C^{I_1 \times I_2 \times \cdots \times I_N}$ and the first mode of tensor $\ten{y} \in \C^{J_1 \times J_2 \times \cdots \times J_M},$ where $I_N = J_1 = K,$ yields an $(N+M-2)$-order tensor $\ten{z}=\ten{x} \bullet \ten{y},$ the entries of which are given by 
\(
z_{i_1 \ldots i_{N-1} j_2 \ldots j_M}=\sum_{k=1}^{K} x_{i_1 \ldots i_{N-1} k} y_{k j_2 \ldots j_M}.
\)
Similarly, a contraction along mode \( n \) is represented by  \( \ten{z} = \ten{x} \bullet_n \ten{y} \), where \( I_n = J_n \). Contraction of all common modes is denoted by  \( \ten{x} \bar{\bullet} \ten{y} \). For tensors of the same size, this is equivalent to the inner product \( \langle \ten{x}, \ten{y} \rangle = \ten{x} \bar{\bullet} \ten{y} \). We use tensor network diagrams to visualize tensor networks, see Fig.~\ref{fignotation}.

\vspace{-3mm}
\vspace{-3mm}
A \emph{qubit} (quantum bit) is the fundamental unit of quantum information, analogous to a classical bit, but with unique quantum properties. Unlike a classical bit, which can be either 0 or 1, a qubit can exist in a superposition of both states simultaneously. Quantum states can be represented as vectors in a Hilbert space, e.g., for a single qubit, the quantum state is a vector in a two-dimensional Hilbert space. The \emph{Pauli spin matrices} for a single qubit are denoted by $\mat{I}_2, \bm{\sigma}_x, \bm{\sigma}_y, \bm{\sigma}_z \in \C^{2 \times 2}$, and are defined by \( \begin{bmatrix} 1 & 0 \\ 0 & 1 \end{bmatrix} \), \( \begin{bmatrix} 0 & 1 \\ 1 & 0 \end{bmatrix} \), \( \begin{bmatrix} 0 & -i \\ i & 0 \end{bmatrix} \) and \(\begin{bmatrix} 1 & 0 \\ 0 & -1 \end{bmatrix} \), respectively. They are used to measure the state of a quantum system.

\begin{figure}[h] 
\centering
\vspace{-3mm}
\begin{tikzpicture}[bluenode/.style={circle, draw=black!100, fill=blue!30, very thick, minimum size=2mm}, scale=0.7]
		% Define coordinates for the nodes
		\foreach \i in {1,2,3,4,5} {
			\node[bluenode] (G\i) at (\i*1.3,0) {};
		}
		
		% Draw the horizontal edges (R_i)
		\foreach \i/\j in {1/2, 2/3} {
			\draw[ultra thick] (G\i) -- node[above] {$R_\i$} (G\j);
		}
		
		\draw[ultra thick,] (G3) -- node[above] {$\cdots$} (G4);
		\draw[ultra thick,] (G4) -- node[above] {$R_{N-1}$} (G5);
        
		% Draw the vertical edges (I_i and J_i)
         \draw[ultra thick] (G1) -- +(0,-.8) node[below] {$I_1$};
         \draw[ultra thick] (G2) -- +(0,-.8) node[below] {$I_2$};
         \draw[ultra thick] (G3) -- +(0,-.8) node[below] {$I_3$};
         \draw[ultra thick] (G4) -- +(0,-.8) node[below]{$I_{N-1}$};
         \draw[ultra thick] (G5) -- +(0,-.8) node[below] {$I_{N}$};

         \draw[ultra thick] (G1) -- +(0,.8) node[above]{$J_1$};
         \draw[ultra thick] (G2) -- +(0,.8) node[above]{$J_2$};
         \draw[ultra thick] (G3) -- +(0,.8) node[above] {$J_3$};
         \draw[ultra thick] (G4) -- +(0,.8) node[above]{$J_{N-1}$};
         \draw[ultra thick] (G5) -- +(0,.8) node[above]{$J_{N}$};
        
\end{tikzpicture}
\vspace{-2mm}
\caption{\small Tensor diagram of a TTM decomposition.}
\label{figTTM}
\end{figure}
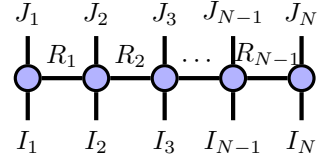

A \emph{tensor train matrix (TTM)}, mathematically equivalent to the MPO, is a special type of tensor train decomposition used to represent a large-scale matrix,  $\mat{X} \in \C^{I \times J},$  as a 2$N$th-order tensor, $\ten{x} \in \C^{I_1 \times J_1 \times I_2 \times J_2 \times \cdots \times I_N \times J_N},$ where $I = I_1 I_2\ldots I_N$ and $J = J_1 J_2\ldots J_N$ \cite{verstraete2007MPS, verstraete2004MPO, oseledets2010tensortrain, oseledets2010approximation}.  The TTM decomposition of $\ten{x}$ is written as a contraction of a sequence of fourth-order core tensors $\ten{g}^{(n)} \in \C^{R_{n-1} \times I_n \times J_n  \times R_{n}} (1 \leq n \leq N)$ with $R_{0} = R_{N} =1$ such that each entry of $\ten{x}$ can be expressed as the sequence of matrix products:
 
 \begin{equation}
 	\label{eqn:ttm1}
 	x_{i_{1} j_{1} i_{2} j_{2} \cdots i_{N} j_{N}} = \mat{g}^{(1)}_{: i_{1} j_{1} :} \mat{g}^{(2)}_{: i_{2}j_{2} :}  \cdots  \mat{g}^{(N)}_{: i_{N}j_{N} :},
 \end{equation} 
 where  matrices $\mat{g}^{(n)}_{: ij :} = \ten{g}^{(n)}(:,i,j,:) \in \C^{R_{n-1} \times R_{n}}$ are slices of the core tensor $\ten{g}^{(n)}.$  The minimal $(R_0, R_1, \ldots, R_{N})$ for which Eq.~\eqref{eqn:ttm1} holds is the TT-rank of $\ten{x},$ denoted by $\vec{r}_{\mathrm{TT}}(\ten{x}).$ This representation can be much more memory-efficient because it only requires storing the core tensors. The TTM decomposition of a given tensor $\ten{x}$ can be computed by a sequence of truncated SVDs (TT-SVD) \cite{oseledets2010tensortrain, oseledets2010approximation}. We denote the (TT-SVD) operation that projects a given tensor $\ten{x}$ to a TTM format, $\left\{\ten{g}^{(n)} \in \C^{R_{n-1} \times I_n \times J_n \times R_{n}}\right\}_{n=1}^N$, by $\ten{P}_{\vec{r}_{\mathrm{TT}}}$. Without loss of generality, we assume that this operation yields a left-orthogonal TT\footnote{All the TT cores to the left of $\ten{g}^{(N)}$ are left-orthogonal \cite{oseledets2010tensortrain, steinlechner2016tc}.}.
 A visualization of the TTM decomposition is shown in Fig.~\ref{figTTM}.
%% TTM diagram
\vspace{-3mm}
If all TT-ranks are equal to 1, then the TTM decomposition is the outer product of $N$ matrices. An $N$-qubit Pauli matrix can be written as the Kronecker product of $N$ Pauli matrices: $\mat{E}_m := \otimes_{j=1}^{N} \bm{\sigma}_{m_j}$ with $\bm{\sigma}_{m_j} \in \{ \mat{I}_2, \bm{\sigma}_x, \bm{\sigma}_y, \bm{\sigma}_z\}$. Such a matrix can be represented in the TTM format by replacing the Kronecker product with the outer product $\ten{E}_m := \op_{j=1}^{N} \bm{\sigma}_{m_j}$ and reversing the order. \par 
The Block-$n$-TT format is a special case of the TTM format, typically used to represent tall matrices, with third-order core tensors except for one fourth-order core at position $n$ \cite{dolgov2014computation, lee2015btt}.
\subsection{Organization}
Section~\ref{sec:qst} provides a concise overview of compressed sensing QST. Next, in Section~\ref{sec:ttqst}, we introduce our approach to TT QST, together with the algorithm detailed in Section~\ref{sec:alg}. Finally, we present numerical experiments in Section~\ref{sec:exp} and a brief conclusion in Section~\ref{sec:con}.

\section{Compressed sensing quantum state tomography}
\label{sec:qst}
Consider an $N$-qubit quantum system with a density matrix $\rho \in \C^{D \times D},$ where $D=2^N.$ According to the quantum theory, when we measure an observable $\mat{E}_m \in \C^{D \times D}$, which is a Hermitian matrix, the measurement outcome is one of its (real) eigenvalues. However, we do not know which eigenvalue from the spectrum will pop up. The determinism in quantum mechanics is only of a statistical character, except for certain laws \cite{nielsen2010quantum}. The expected measurement outcome can be determined and is given by $y_m = \operatorname{Tr}(\rho \mat{E}_m).$ QST estimates the unknown density matrix by performing measurements on an ensemble of identically prepared quantum systems (i.e., prepared in the same quantum state). In estimation terminology, given (possibly noisy) measurement data $\{\mathbf{E}_m, y_m\}_{m=1}^{M},$ we are required to estimate a density matrix $\hat{\rho}$ that agrees with the observed data, i.e., $y_m \approx \operatorname{Tr}(\hat{\rho} \mat{E}_m), \text{ for } m=1, 2, \ldots, M$. The number of measurements required to ensure the unique recovery of the (full-rank) density matrix is approximately $\mathcal{O}(D^2)$.  However, in the rank-$R$ case, \(\mathcal{O}(RD\log^2(D))\) measurements suffice to uniquely recover the density matrix with high probability \cite{gross2010csqst, liu2011universal, liu2012csqst, wang2013qstvmc, candes2012mc, kalev2015quantum}. The rank minimization of QST (with nuclear norm relaxation of the rank) can be expressed as:
\begin{equation}
\label{eqn:sdpQST}
\begin{aligned}
& \min_{\hat{\rho}} \|\hat{\rho}\|_{*} \quad \text{s.t.} \quad \|\vec{y} - \ten{M}(\hat{\rho})\|_2 \leq \epsilon, \text{ and } \hat{\rho} \in \mathcal{S},  \\
& \text{where } \mathcal{S} = \{ \mat{x} \mid  \mat{x} \succeq 0, ~ \operatorname{Tr}(\mat{x})=1, ~  \mat{x} = \mat{x}^{\mathrm{H}} \},
\end{aligned}
\end{equation}
where $\vec{y} \in \R^M$ is the vector of measurement outcomes and the measurement map $\mathcal{M}$ is defined such that $\left(\mathcal{M}(\mat{x})\right)_m = \operatorname{Tr}(\mathbf{E}_m \mat{x})$.  When $N$ is small, the problem \eqref{eqn:sdpQST} can be easily solved using software such as SDPT3 or  CVX(PY) \cite{cvxpy}. For large-scale systems, optimizing over a full density matrix becomes expensive. Alternative approaches have been explored that explicitly impose a low-rank decomposition model and optimize over low-rank factors. E.g., it has been shown that one can recast the problem \eqref{eqn:sdpQST} into a problem that implicitly accounts for the Hermitian and PSD constraints by setting $ \hat{\rho} = \mat{A}\mat{A}^{\mathrm{H}}, $ where $\mat{A} \in \mathbb{C}^{D \times R}$; see, e.g., \cite{burer2003lrsdp}. In general, the SDP of this parametrization can have local solutions\footnote{Without any additional constraints beyond the SDP constraint, one can decompose $\rho$ as $\rho  = \mat{\hat{A}} \mat{\hat{A}}^{\mathrm{H}},$ where $\mat{\hat{A} = AL}$ for any unitary $\mat{L} \in \C^{R \times R}$ such that $\mat{LL}^{\mathrm{H}} = \mat{I}$.}. However, the extra unit trace constraint makes it suitable for many tasks, such as QST \cite{bhojanapalli2016lrsdp, burer2003lrsdp, kalev2015quantum, kyrillidis2018provable, kim2023mifgd}. More recently, projected (factored) gradient descent methods have been proposed with further relaxation of the unit trace by a convex constraint $\|\mat{A}\|_{\mathrm{F}}^2\leq 1 \Leftrightarrow \operatorname{Tr}( \mat{\hat{\rho}}) \leq1$ \cite{kalev2015quantum,kyrillidis2018provable}. Our approach generalizes such parametrizations to tensor networks.

\section{Tensor train quantum state tomography}
\label{sec:ttqst}
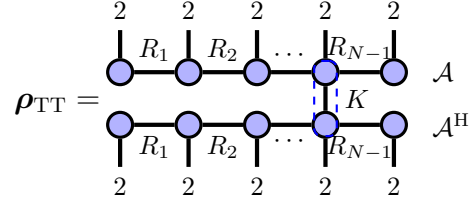
\begin{figure}[h]
	\centering
\begin{tikzpicture}[bluenode/.style={circle, draw=black!100, fill=blue!30, very thick, minimum size=1mm}, scale=0.7]
% First figure with scope and xshift
\begin{scope}[xshift=0, yshift=0.4cm]
	\node at (0,0) {\large $~ \rho_{\mathrm{TT}} = $};
	\end{scope}
		% Second figure with scope and xshift
	\begin{scope}[xshift=0cm, yshift=1cm] % Shift the second figure to the right
		% Define coordinates for the nodes in the top diagram
		\foreach \i in {1,2,3,4,5} {
			\node[bluenode] (G\i) at (\i*1.3,0) {};
		}
		
		% Draw the horizontal edges (R_i) for the top diagram
		\foreach \i/\j in {1/2, 2/3} {
			\draw[ultra thick] (G\i) -- node[above] {$R_\i$} (G\j);
		}
		
		\draw[ultra thick,] (G3) -- node[above] {$\cdots$} (G4);
		\draw[ultra thick,] (G4) -- node[above] {$R_{N-1}$} (G5);
		
		% Draw the vertical edges (I_i) for the top diagram
		\foreach \i in {1,2,3,4,5} {
			\draw[ultra thick] (G\i) -- +(0, .8) node[above] {$2$};
		}

		% Define coordinates for the nodes in the bottom diagram (upside down)
		\foreach \i in {1,2,3,4,5} {
			\node[bluenode] (H\i) at (\i*1.3,-1) {};
		}
		
		% Draw the horizontal edges (R_i) for the bottom diagram
		\foreach \i/\j in {1/2, 2/3} {
			\draw[ultra thick] (H\i) -- node[below] {$R_\i$} (H\j);
		}
		
		\draw[ultra thick,] (H3) -- node[below] {$\cdots$} (H4);
		\draw[ultra thick,] (H4) -- node[below] {$ R_{N-1}$} (H5);
		
		% Draw the vertical edges (I_i) for the bottom diagram
		\foreach \i in {1,2,3,4,5} {
			\draw[ultra thick] (H\i) -- +(0,-.8) node[below] {$2$};
		}
		
		% Connect both diagrams at node J_alpha
		\draw[ultra thick,] (G4) -- (H4);
		
		\node[draw, blue, thick, dashed, rounded corners=.1cm, minimum width=.3cm, minimum height=1cm] at (5.2, -0.5) {};
        \node at (5.8, -0.5){$K$};
\node[right = 0.2 of G5]{$\ten{a}$};
\node[right = 0.2 of H5]{$\ten{a}^{\mathrm{H}}$};
\end{scope}
	
\end{tikzpicture}
\vspace{-2mm}
	\caption{\small Decomposition of $\rho_{\mathrm{TT}}$ into Block-TT format, i.e.,  $\rho_{\mathrm{TT}} = \ten{A}\bullet_{(N-1)} \ten{A}^{\mathrm{H}},$ where $\ten{A}  \in \mathbb{C}^{2 \times 2 \times \cdots  \times K \times 2}.$}
    \vspace{-3mm}
\label{figTMM2BTT}
\end{figure}

\begin{figure}[h]
	\centering
\begin{tikzpicture}[bluenode/.style={circle, draw=black!100, fill=blue!30, very thick, minimum size=2mm}, scale=0.7]
% First figure with scope and xshift
\begin{scope}[xshift=-0.5cm, yshift=0cm]
	\node at (0,0) {\large $y_m = \rho_{\mathrm{TT}} \bar{\bullet}\ten{E}_m = $};
	\end{scope}
		% Second figure with scope and xshift
	\begin{scope}[xshift=1.3cm, yshift=1cm] % Shift the second figure to the right
		% Define coordinates for the nodes in the top diagram
		\foreach \i in {1,2,3,4,5} {
			\node[bluenode] (G\i) at (\i*1,0) {};
		}
		
		% Draw the horizontal edges (R_i) for the top diagram
		\foreach \i/\j in {1/2, 2/3} {
			\draw[ultra thick] (G\i) -- node[above] {} (G\j);
		}
		
		\draw[ultra thick,] (G3) -- node[above] {$\cdots$} (G4);
		\draw[ultra thick,] (G4) -- node[above] {} (G5);

		% Define coordinates for the nodes in the bottom diagram (upside down)
		\foreach \i in {1,2,3,4,5} {
			\node[bluenode] (H\i) at (\i*1,-1) {};
		}
		
		% Draw the horizontal edges (R_i) for the bottom diagram
		\foreach \i/\j in {1/2, 2/3} {
			\draw[ultra thick] (H\i) -- node[below] {} (H\j);
		}
		
		\draw[ultra thick,] (H3) -- node[below] {$\cdots$} (H4);
		\draw[ultra thick,] (H4) -- node[below] {} (H5);
		
		% Draw the vertical edges (I_i) for the bottom diagram
		\foreach \i in {1,2,3,4,5} {
			\draw[ultra thick] (H\i) -- +(0,-.8) node[below] {};
		}
		
		% Connect both diagrams at node J_alpha
		\draw[ultra thick,] (G4) -- (H4);
        \node at (4.5, -0.5){};
\node[right = 0.2 of G5]{$\ten{a}$};
\node[right = 0.2 of H5]{$\ten{a}^{\mathrm{H}}$};

% Pauli operators
		% Define coordinates for the nodes in the bottom diagram (upside down)
		\foreach \i in {1,2,3,4,5} {
			\node[bluenode] (E\i) at (\i*1,-1.7) {};
		}
		
		% Draw the horizontal edges (R_i) for the bottom diagram
		\foreach \i/\j in {1/2, 2/3} {
			\draw[draw=none] (E\i) -- node[below] {} (E\j);
		}
		
		\draw[draw=none] (E3) -- node[below] {} (E4);
		\draw[draw=none] (E4) -- node[below] {} (E5);
		
		% Draw the vertical edges (I_i) for the bottom diagram
		\foreach \i in {1,2,3,4,5} {
			\draw[ultra thick, black!60] (G\i) to[out=115,in=-115, looseness=1.8] (E\i);
		}
		
		% Connect both diagrams 
         \foreach \i in {1,2,3,4,5} {
			\draw[ultra thick] (E\i) -- (H\i);
		}

\node[draw,  thick, blue,  dashed, rounded corners=.1cm, minimum width=3.3cm, minimum height=0.9cm] at (3, -1.35) {};        
\node[right = 0.2 of E5]{$\ten{E}_m$};

\end{scope}
\end{tikzpicture}
\vspace{-7mm}
\caption{\small Tensor network contraction for computing the expectation value $\rho_{\mathrm{TT}} \bar{\bullet}\ten{E}_m,$ with $\rho_{\mathrm{TT}}$ and $\ten{E}_m$ represented in TT format.}  
%Expectation values are efficiently calculated by a tensor network contraction. First, \(\ten{A}^{\mathrm{H}}\) is contracted with \(\ten{E}_m\) along common modes (shown in the area in the dashed line, i.e.,   \(\ten{A}^{\mathrm{H}} \bar{\bullet} \ten{E}_m\)), and then the resulting tensor is contracted with \(\ten{A}\) along all the modes.
\label{figDensityTrace}
\vspace{-2mm}
\end{figure}
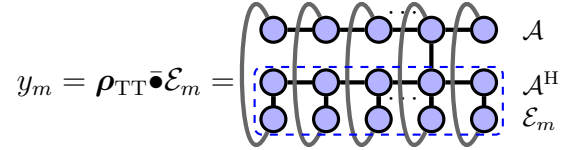

As noted earlier, the MPS (TT) efficiently represents a range of quantum many-body (pure) states and comes with more advanced algorithms, known as density matrix renormalization group algorithms (DMRG), for solving one-dimensional quantum systems \cite{schollwock2005dmrg, oseledets2011dmrg, verstraete2007MPS}. The MPO (TTM) extend the MPS  to mixed states \cite{verstraete2004MPO}. In this work, we employ a representation that expresses a density matrix as a contraction of two Block-TT networks.  This parametrization allows for representing a mixed state with asymptotically the same number of parameters as that of a TT\footnote{The number of parameters is equal to  $2R^2 (\log_2D-3) +  2R^2 K + 4R $, assuming $R_n=R$.}. Moreover, this representation inherently ensures the positive semi-definiteness of the corresponding density matrix by construction. In this representation, a \(2N\)th-order density tensor \(\rho \in \C^{2 \times 2 \times \cdots \times 2}\) is expressed as a contraction of an \((N+1)\)th-order Block-TT \(\ten{A}\) with its Hermitian transpose \(\ten{A}^{\mathrm{H}}\), i.e., $\rho_{\mathrm{TT}} = \ten{A}\bullet_n \ten{A}^{\mathrm{H}}$. Intuitively, this is similar to the Burer-Monteiro factorization for matrices \cite{burer2003lrsdp}.  A tensor network diagram of this model is shown in Fig.~\ref{figTMM2BTT}. It is interesting to note that the position of the 4th-order core carrying the index $K$ can be moved to other locations. It has been shown that it serves as an extra tuning parameter during optimization \cite{dolgov2014computation, lee2015btt}. Furthermore, this can be utilized to adjust ranks dynamically: the 4th-order core can be merged with one of its neighbours, and subsequently, this combined core can be decomposed using an SVD of a different rank, which also allows the index $K$ to be shifted to an adjacent position.
 \vspace{-5mm}
\unskip
Although there are different choices for observables (measurements), we use $N$-qubit Pauli matrices in the TTM format $\{ \ten{E}_m\}_{m=1}^M$, as they can be decomposed with TT-ranks equal to 1. We utilize the TT algebra to efficiently compute expectation values through tensor network contraction, as shown in Fig.~\ref{figDensityTrace}, which represents the inner product between $\rho_{\mathrm{TT}}$ and $\ten{E}_m$. 
\vspace{-10mm}
\unskip
 First, \(\ten{A}^{\mathrm{H}}\) is contracted with \(\ten{E}_m\) along common modes i.e.,  \(\ten{A}^{\mathrm{H}} \bar{\bullet} \ten{E}_m\) (shown in the area in the dashed line). The number of arithmetic operations required for this step is \(\mathcal{O}\bigl(4 R^2 (\log_2D + K)\bigr)\), assuming $R_n = R$. Subsequently, the resulting tensor is contracted with \(\ten{A}\) along all modes. The complexity of this step is \(\mathcal{O}\bigl(2 R^3 (\log_2D + K)\bigr)\). Therefore, the total computational complexity is \(\mathcal{O}\bigl((4 R^2 + 2 R^3)(\log_2D + K)\bigr)\); see \cite{oseledets2010tensortrain, oseledets2011dmrg,  lee2018tenops} for more details. If the TT-rank and $K$ are small, this approach can be much more efficient than the standard matrix trace computation $\operatorname{Tr}(\rho\,\mat{E}_m)$, which requires $\mathcal{O}(D^2)$ arithmetic operations when computed efficiently.

\subsection{Optimization objective} \label{sec:opt}
We recast the problem~\eqref{eqn:sdpQST}, in line with \cite{bhojanapalli2016lrsdp, burer2003lrsdp, kalev2015quantum, kyrillidis2018provable, kim2023mifgd}, as a non-convex problem: 
\begin{equation}
		\label{eqn:pfgd}
		\begin{array}{ll}
			\displaystyle \min_{\ten{A}} & \underbrace{\frac{1}{2} \| \vec{y}  - \ten{M}\left(\hat{\rho}_{\text{TT}}\right) \|_2^2}_{ =: f\left(\ten{A} \right)}, ~ \text { with } \hat{\rho}_{\mathrm{TT}} = \ten{A}\bullet_n \ten{A}^{\mathrm{H}}.\\
			\text { s.t. } & \|\ten{A}\|_{\mathrm{F}}^2\leq 1,
		\end{array}
\end{equation} 
where the entries of $\ten{A}$ are given by \(a_{i_{1} i_{2} \cdots i_{n}  k_{n} \cdots  i_{N} } = \mat{g}^{(1)}_{: i_{1}:} \mat{g}^{(2)}_{: i_{2} :}  \cdots  \mat{g}^{(n)}_{: i_{n}, k_{n} :} \cdots \mat{g}^{(N)}_{: i_{N} :}\).
We have reformulated the problem for two reasons: firstly, the number of parameters in the proposed model scales logarithmically with the size of the density matrix\footnotemark[3]. Secondly, this allows us to utilize tensor network operations to perform computations efficiently. It is important to note that this approach is efficient only when the quantum states can be well approximated with low TT-rank. The gradient of the objective \eqref{eqn:pfgd}, given by $\nabla f\left(\ten{A} \right) = -2 \sum_{m=1}^{M} \left(y_m - \hat{\rho}_{\mathrm{TT}} \bar{\bullet} \ten{E}_m\right) \ten{E}_m \bar{\bullet} \ten{A},$ can be computed core-wise.  
\subsection{Algorithm} \label{sec:alg}
This section provides a simple gradient descent (GD) algorithm for optimizing the objective~\eqref{eqn:pfgd}. The key steps of the algorithm are as follows. A TT-SVD operation $\ten{P}_{\vec{r}_{\mathrm{TT}}} \left(\ten{A} \right)$ is performed, which projects $\ten{A}$ to a left-orthogonalized Block-TT format;  see \cite[Algorithm~1 and Algorithm~2]{oseledets2010tensortrain} for more information. As noted earlier, this process involves truncation and orthonormalization. Truncation (or rounding) is required because the TT-ranks of the gradient and the current iterate get added following the GD step. Alternative approaches that optimize over fixed-rank TT manifolds (see \cite{holtz2012manifolds, lubich2015time, steinlechner2016tc, vermeylen2025riemannian}) could be more suitable; however, here we provide a simple working algorithm. As $\ten{A}$ is left-orthogonal, the Frobenius norm of $\ten{A}$ is equal to the norm of its last core, $\ten{g}^{(N)}$. In the projection step, we normalize $\ten{g}^{(N)}$ to ensure that $\|\ten{A}\|_{\mathrm{F}}^2\leq 1$. We noticed that adding a small momentum factor $\eta$ in the GD step improved the convergence. The pseudocode for the proposed method is outlined in Algorithm~\ref{alg:ttpfgd}.
\begin{algorithm}[htb]
\caption{Block-TT QST Algorithm.}\label{alg:ttpfgd}
\KwData{Gradient of objective $ \nabla f\left(\ten{A} \right)$,  $\eta$,  $\vec{r}_{\mathrm{TT}}$.\\}
\KwResult{$\ten{A},$ such that $\hat{\rho}_{\mathrm{TT}} =  \ten{A}\bullet_{(N-1)} \ten{A}^{\mathrm{H}}$.}
Initialize $\ten{A}_i$.  \\
$\ten{A}_i = \ten{P}_{\vec{r}_{\mathrm{TT}}} \left(\ten{A}_i \right).$   \textcolor{gray}{\% $\ten{A}_i$ is left-orthogonal} \\
Projection: $ \ten{A}_i = \frac{1}{\|\ten{g}^{(N)}\|_{\mathrm{F}}}\ten{A}_i$. ~   \textcolor{gray}{\% $ \|\ten{A}\|_{\mathrm{F}}^2\leq 1$}\\
\For{$ i = 1 \ldots$ }{
Gradient Descent:  $\ten{A}_{i+1} = \ten{P}_{\vec{r}_{\mathrm{TT}}} \left(\ten{A}_i - \eta \nabla f\left(\ten{A}_i \right)\right).$\\
Projection: $ \ten{A}_{i+1} = \frac{1}{\|\ten{g}^{(N)}\|_{\mathrm{F}}}\ten{A}_{i+1}$.\\
}
\end{algorithm}
\vspace{-5mm}
\section{Experiments}
\label{sec:exp}

Experiments are conducted on an HP EliteBook 845 G8 with an AMD Ryzen 7 PRO 5850U processor and 32GB RAM. For tensor-train computations, the open-source Python package \texttt{Scikit-TT}\footnote{\url{https://github.com/PGelss/scikit_tt}} is used.
\vspace{-2mm}
\subsection{Evaluation Metrics} \label{sec:metrics}
Two common metrics are used to assess performance: (1) Fidelity, which quantifies the closeness between two density matrices \(\rho_1\) and \(\rho_2\): \( \left( \operatorname{Tr}\left(\sqrt{\sqrt{\rho_1} \rho_2 \sqrt{\rho_1}} \right) \right)^2 \) and, (2) Trace distance, which quantifies distinguishability: \(\frac{1}{2} \|\rho_1 - \rho_2\|_1\). 
\subsection{Results} \label{sec:results}
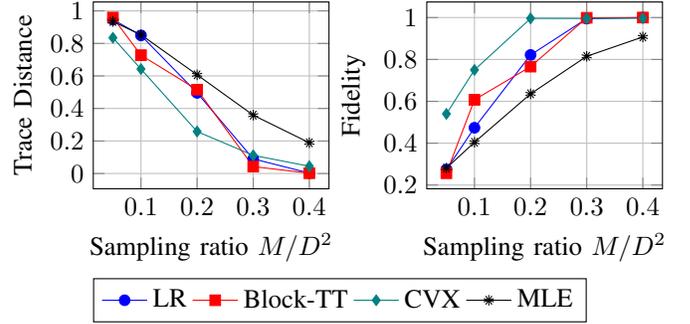
\begin{figure}[h] 
\centering
\pgfplotsset{compat=newest}
  \begin{tikzpicture}
    \begin{groupplot}[
      group style={group size=2 by 1, horizontal sep=1.3cm},
      width=0.26\textwidth,
      legend to name=sharedlegend,
      xlabel={Sampling ratio $M/D^2$}, legend columns=-1
    ]
      \nextgroupplot[
        title={},
        domain=0:5, samples=100,
        ylabel={Trace Distance},
        tick label style={font=\normalsize},
         ytick={0.0,0.2,0.4,0.6,0.8,1},
        grid=major,
        every axis label/.append style={font=\normalsize},  
        every axis tick label/.append style={font=\normalsize},
      ]
        \addplot[color=blue, mark=*] coordinates {(0.05, 0.94709727) (0.10, 0.84912094) (0.20, 0.49585407) (0.30, 0.08986963) (0.40, 0.00271078)};
        \addlegendentry{LR}
        
        \addplot[color=red, mark=square*, ] coordinates {(0.05, 0.95905674) (0.10, 0.7277381) (0.20, 0.51554798) (0.30, 0.04259094) (0.40, 0.00172938)};
        \addlegendentry{Block-TT}
        
        \addplot[color=teal, mark=diamond*] coordinates {(0.05, 0.83482417) (0.10, 0.64220898) (0.20, 0.25811959) (0.30, 0.11180522) (0.40, 0.04447706)};
        \addlegendentry{CVX}
        
        \addplot[color=black, mark=10-pointed star, ] coordinates {(0.05, 0.93486451) (0.10, 0.85320154) (0.20, 0.60698464) (0.30, 0.35874785) (0.40, 0.18899541)};
        \addlegendentry{MLE}
      
      % Fidelity plot
      \nextgroupplot[
        title={},
        domain=0:5, samples=100,
        ylabel={Fidelity},
        tick label style={font=\normalsize},
        ytick={0,0.2,0.4,0.6,0.8,1},
        grid=major,
        every axis label/.append style={font=\normalsize},  
        every axis tick label/.append style={font=\normalsize},
      ]
       \addplot[color=blue, mark=*] coordinates {(0.05, 0.2742909) (0.10, 0.47361944) (0.20, 0.82152764) (0.30, 0.99496474) (0.40, 0.99999619)};
        \addlegendentry{LR}
        
        \addplot[color=red, mark=square*, ] coordinates {(0.05, 0.25650167) (0.10, 0.60720161) (0.20, 0.76564993) (0.30, 0.99886428) (0.40, 1.000013)};
        \addlegendentry{Block-TT}
        
        \addplot[color=teal, mark=diamond*] coordinates {(0.05, 0.53996843) (0.10, 0.74985683) (0.20, 0.995971) (0.30, 0.99560987) (0.40, 0.99655413)};
        \addlegendentry{CVX}
        
        \addplot[color=black, mark=10-pointed star, ] coordinates {(0.05, 0.28079222) (0.10, 0.4042542) (0.20, 0.6355638) (0.30, 0.81489767) (0.40, 0.90735691)};
        \addlegendentry{MLE}
      
    \end{groupplot}
    
    \node at ($(current bounding box.south)+(0,-0.4cm)$) {\pgfplotslegendfromname{sharedlegend}};
  \end{tikzpicture}
\vspace{-5mm}
\caption{\small The proposed method outperforms MLE and has an overall similar performance to LR. The CVX method performs best at low sampling ratios ($M/D^2 < 0.3$), but this approach solves the original problem~\eqref{eqn:sdpQST} and becomes very expensive as the problem size increases.}
\label{fig:comp}
\vspace{-4mm}
\end{figure}

An $8$th-order tensor in Block-TT format $\ten{A} \in \C^{2 \times 2 \times \cdots \times K \times 2}$ is generated with $K = 2$ and $\vec{r}_{\mathrm{TT}}(\ten{A}) = (1, 2, \ldots, 2, 1)$, by sampling entries of the core tensors from a normal distribution (the Ginibre ensemble). The tensor is normalized and a $14$th-order tensor $\rho_{\mathrm{TT}} = \ten{A} \bullet_{(N-1)} \ten{A}^{\mathrm{H}}$ is computed. Random Pauli measurements are performed with sampling ratio $M/D^2$ varying from $0.05$ to $0.4$. Gaussian noise is added to the measurements with $\mathrm{SNR} = \qty{60}{\decibel}$. Algorithm \ref{alg:ttpfgd} is compared with state-of-the-art algorithms in the reconstruction of $\hat{\rho}_{\mathrm{TT}}$, including a convex optimization method (CVX) \cite{gross2010csqst, cvxpy}, maximum likelihood estimation (MLE) \cite{banaszek1999maximum}, and a low-rank projected factor gradient method (LR) \cite{kyrillidis2018provable}. The LR algorithm was initialized randomly. The median Fidelity and Trace distance are shown across 20 trials in Fig.~\ref{fig:comp}. 

% Next, we generated \(2N\)th-order tensors \(\rho_{\mathrm{TT}}\) similar to those from the previous experiment. We varied \(N\) from \(3\) to \(5\). We computed the reconstruction \(\hat{\rho_{\mathrm{TT}}}\) using the proposed method and recorded the trace and eigenvalues of the corresponding density matrix. The distribution of the trace and eigenvalues is plotted across 1000 trials in Fig.~\ref{fig:a}. It can be seen that the reconstructions are indeed valid in that they satisfy the constraints of being a density matrix.

% \begin{figure}[h]
%     \centering
%     \subfigure[]{
%         \includegraphics[width=0.23\textwidth, trim={2 2 2 2},clip]{figures/ttqst34valid}
%         \label{fig:a}
%     }
%     \hspace{-5mm}
%     \subfigure[]{
%         \includegraphics[width=0.23\textwidth, trim={2 2 2 2},clip]{figures/ttqsttime}
%         \label{fig:b}
%     }
%     \caption{\small{(a) Eigenvalues lie on the real line between 0 and 1, satisfying Hermitian and positive semi-definite (PSD) constraints. We can also see that the trace of the reconstructions is one; (b) The time required to perform the measurement in the Block-TT format scales close to linearly, whereas in a matrix format, it scales exponentially.}}
%     \label{fig:sbs}
% \end{figure}

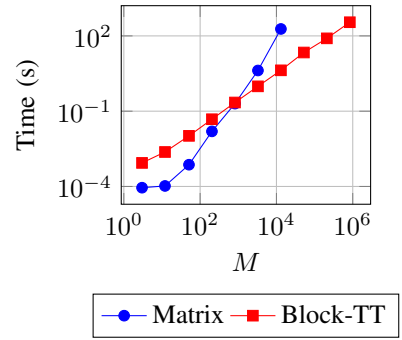
\begin{figure}[h]
  \centering
      \begin{tikzpicture}
    \begin{axis}[
      width=0.27\textwidth,
      xlabel={$M$},
      xtick={1,10^2,10^4,10^6},
      ylabel = {Time (s)},
      legend style={at={(0.5,-0.45)}, anchor=north, legend columns=2},  
      grid=both,
      ymode=log,
      xmode=log
    ]
      
      \addplot[color=blue, mark=*] coordinates {
        (3, 8.929998148232698e-05)
        (12, 0.00010480009950697422)
        (51, 0.000735800014808774)
        (204, 0.01568000006955117)
        (819, 0.2002502999966964)
        (3276, 4.18418609991204)
        (13107, 189.22)
      };
      \addlegendentry{Matrix}

      \addplot[color=red, mark=square*, ] coordinates {
        (3, 0.0008675999706611037)
        (12, 0.002364400075748563)
        (51, 0.010385300032794476)
        (204, 0.0478129000402987)
        (819, 0.22202380001544952)
        (3276, 0.9770273000467569)
        (13107, 4.252329399925657)
        (52428, 21.801936500007287)
        (209715, 79.9286891000811)
        (838860, 350.7832464000676)
      };
      \addlegendentry{Block-TT}
    \end{axis}
  \end{tikzpicture}
\caption{ \small The time required to perform $M$ measurements in the Block-TT format scales close to linearly, whereas in a matrix format, it scales exponentially.}
\label{fig:time}
\vspace{-5mm}
\end{figure}

Next, the number of qubits \(N\) is varied from \(3\) to \(12\). Random Pauli measurements are performed,  with the sampling ratio \(M/D^2\) set to \(0.05\). Note that, as \(N\) increases, \(D\) increases proportionally to \(2^N\) and  \(M\) increases proportionally to \(D^2\) (i.e., \(4^N\)). Fig.~\ref{fig:time} shows the median time required to perform $M$ measurements, i.e., \(\{\rho_{\mathrm{TT}} \bar{\bullet} \ten{E}_m\}_{m=1}^M\), across 10 trials. This is the most computationally expensive step in evaluating the cost function and its gradient. In the matrix case, performing $M$ measurement, i.e., \(\{\operatorname{Tr}\left(\rho \mat{E}_m\right)\}_{m=1}^M\), becomes exceedingly expensive as \(N\) increases, while the proposed format allows performing measurements for large-scale systems. 

\section{Conclusion and future work}
\label{sec:con}
A novel QST approach is introduced in which a density matrix is parameterized by a Block-TT network. The proposed method is both memory- and computationally efficient. Numerical experiments show that the proposed approach yields accurate and valid results without additional constraints. Further theoretical analysis, e.g., of the recovery guarantees, could be useful (some theoretical results of a related approach are known \cite{rauhut2017ttIHT}). More efficient algorithms can be designed, for example, by optimizing over TT manifolds \cite{holtz2012manifolds, lubich2015time, steinlechner2016tc, vermeylen2025riemannian}. Furthermore, structured tensor completion approaches could be utilized to perform QST algebraically with deterministic recovery guarantees; see, e.g., \cite{nico2014curseofdim, sorensen2019fiber, stijn2023mlsvdfsj, shakir2024ttfw}.
\printbibliography
\end{document}